\documentclass{PoS}

\usepackage{multirow}
\usepackage{gensymb}
\def \babar {\mbox{\slshape B\kern-0.1em{\smaller A}\kern-0.1em B\kern-0.1em{\smaller A\kern-0.2em R}}}

\title{Precise measurements of CP violation in B decays at Belle II}

\ShortTitle{Precise measurements of CP violation in B decays at Belle II}

\author{\speaker{Benjamin Oberhof}\thanks{On behalf of the Belle II collaboration.}\\
        Laboratori Nazionali di Frascati dell'INFN\\
        E-mail: \email{benjamin.oberhof@lnf.infn.it}}

\abstract{We report expected sensitivities for the measurement of the three angles of the Unitarity Triangle in $B$ meson decays using 50 ab$^{-1}$ of $e^+ e^-$ collisions recorded by the Belle II detector 
at the SuperKEKB asymmetric $e^+ e^-$ collider at KEK, Japan.}

\FullConference{XIV International Conference on Heavy Quarks and Leptons (HQL2018)\\
		May 27- June 1, 2018\\
		Yamagata Terrsa, Yamagata, Japan}

\begin{document}

\section{Introduction}
In the Standard Model (SM) flavor changing quark interactions are described by the unitary Cabibbo-Kobayashi-Maskawa (CKM) unitary matrix $V_{CKM}$. 
The unitarity condition $V^{ \dag } V = I$ gives nine equations, six of which, corresponding to the off-diagonal elements of the identity matrix $I$, 
can be represented as triangles in the complex plane. 
The most commonly used triangle at B-factories is the one given by the following equation 
\begin{equation}
\sum_i V_{id} V^* _{ib} = 0, \mbox{         } (i=u,c,t). 
\label{ut_def}
\end{equation}
The branching fractions for $B$ decays are proportional to the sides of this triangle while the angles are proportional to the amount of CP-violation (CPV). 
Two different naming conventions are in use for the angles of this triangle $\alpha$, $\beta$ and $\gamma$ and $\phi_1$, $\phi_2$ and $\phi_3$ respectively. 
In the following we will use the latter one. The relations among the two conventions are $\alpha$=$\phi_2$, $\beta$=$\phi_1$ and $\gamma$=$\phi_3$. 
All three angles can be measured at B-factories using different $B$ decay modes \cite{physics}. $\phi_1$ and $\phi_2$ can be measured in time-dependent CP-violation analysis (TDCPV): 
for $\phi_1$ in $b \rightarrow c \bar c s$ and $b \rightarrow q \bar q s$, $q=u,d,s$ transitions, and for $\phi_2$ in $B \rightarrow h h$, $h = \pi, \rho$ decays. 
$\phi_3$ finally can be accessed by the measurement of the interference in $b \rightarrow c \bar u s$ and $b \rightarrow u \bar c s$ amplitudes. 

%\begin{figure*}[hbt!]
%\centering
%\includegraphics[width=0.7\textwidth, angle=0, clip]{ut_b_decays.pdf} 
%\caption{Unitarity triangle defined in \ref{ut_def} and golden modes for the measurement of its angles at B-factories.}
%\label{unit}
%\end{figure*} 

\section{The Belle II experiment}

The Belle II experiment follows the path defined by the Belle and BaBar experiments, both of which started about 20 years ago at the B-factories KEKB (Tsukuba, Japan) and PEP-II (SLAC, USA) respectively. 
Until now all measurements made at B-factories are in agreement with the Standard Model; nowadays, however, there is compelling evidence for New Physics beyond the Standard Model from various sources 
(e.g. neutrino mixing, baryonic asymmetry in the universe). For this reason Japan has decided to upgrade the existing KEKB accelerator to deliver a 40 times higher instantaneous luminosity which will allow, 
in 5 years of data taking, to record a data sample 50 times larger than that recorded, jointly, by BaBar and Belle. 
%\begin{figure*}[htb!]
%\centering
%\includegraphics[width=6.cm, angle=270, clip]{belle2.pdf} 
%\caption{Overview of the SuperKEKB B-factory (left) and Belle II detector (right). In color the new or updated parts of SuperKEKB with respect to KEKB.}
%\label{superkekb}
%\end{figure*} 
The new machine, called SuperKEKB, has now been completed and commissioning has started. The design luminosity is $8 \times 10^{35} $cm$^{-2} s^{-1}$ 
with a projected integrated luminosity of 50 ab$^{-1}$ in 5 years running at the center of mass energy of the $\Upsilon(4S)$. 
Because of the increased level of background, the Belle II detector has to cope with higher occupancy and radiation damage than the Belle detector. 
To be able to operate at the conditions of the SuperKEKB collider, the components of the Belle detector are either upgraded or replaced by new ones. 
A new vertex detector (VXD) is being built, a new drift chamber (CDC) with smaller cell size has been built, the particle identification system will include a new Time Of Propagation (TOP) 
detector. The barrel CsI crystals, thallium doped, EM calorimeter (ECL) will be provided with new readout; studies are being carried on for the end caps upgrade. In the $K_L$ and muon detector (KLM) 
only the outer barrel layers of glass RPCs will be re-used, the remaining will be substituted with scintillation counters. For a detailed description of the detector please refer to \cite{tdr}.
\begin{figure}[hbt!]
\centering
\includegraphics[width=0.5\textwidth, angle=270]{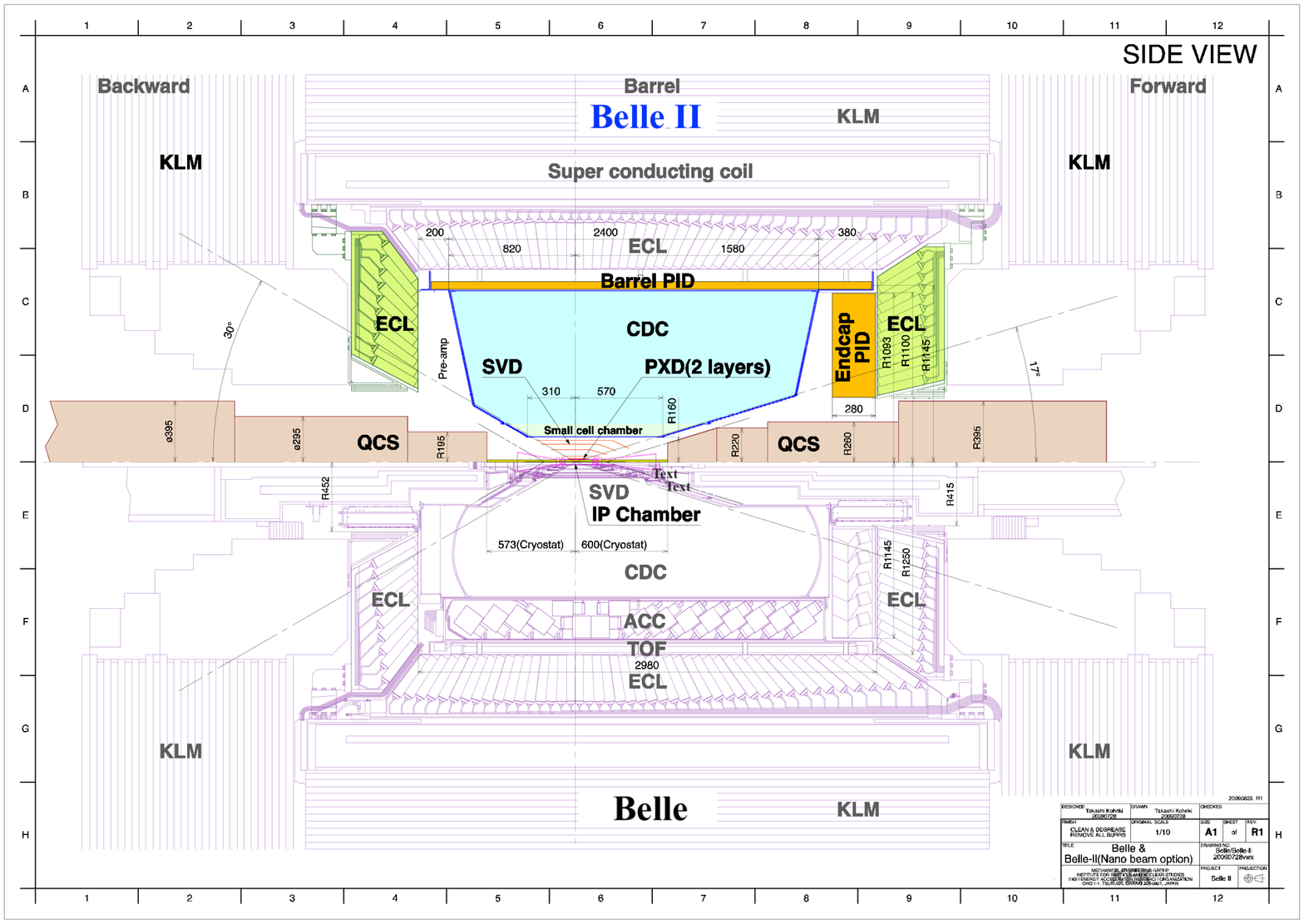} %[width=8.2cm, clip]
\caption{Comparison among the Belle (bottom) and Belle II (top) detectors. In color the updated parts.}
\label{detector}
\end{figure} 
Commissioning (Phase 1) of the main ring (without final focus quadrupoles) has been successfully carried out between 
February and June 2016; instead of Belle II, a commissioning detector, BEAST II (Beam Exorcism for A STable experiment II), was used, in order to measure 
actual beam induced background rates at the Interaction Point (IP). The roll in of the Belle II detector without vertex detector, which will be replaced by a modified version of the BEAST II detector, 
has taken place in Early 2017 followed by a global cosmic run. 
The second phase of commissioning (Phase 2) and first physics runs has started in early 2018; 
during summer 2018 shutdown the vertex detector will be installed and Phase 3 data-taking, with full Belle II detector, is scheduled to start by the end of 2018. 

\section{Measurement of $\phi_1$}
\label{sec_phi1}
The best way to measure $\phi_1$ at B-factories is the TDCPV analysis of $b \rightarrow c \bar c s$ and $b \rightarrow q \bar q s$ transitions. While the tree-level process $b \rightarrow c \bar c s$ results 
in a more precise measurement, especially using the golden mode $B \rightarrow J/\psi K_S$, $b \rightarrow q \bar q s$, being a loop process, is also sensitive to New Physics (NP) contributions. 
The general strategy is to take advantage of the entangled production of the $B^0 \bar B^0$ meson pair trough the reaction $e^+e^- \rightarrow \Upsilon(4S) \rightarrow B^0 \bar B^0$. 
The $B$ meson pair evolves in a coherent state, with exactly one $B^0$ and one $\bar B^0$ at any given time, until one of the mesons decays. 
When one of the mesons decays the other $B$ meson continues to propagate and oscillate between a $B^0$ and $\bar B^0$ state until its own decay. 
%\begin{figure}[htb!]
%\centering
%\includegraphics[width=0.6\textwidth, angle=0]{tdcpv.pdf} %[width=8.2cm, clip]
%\caption{Measurement strategy for TDCPV analysis.}
%\label{tdcpv}
%\end{figure} 
The identification of one of the $B$ mesons, defined as tagging $B$, as a $B^0$ or $\bar B^0$, defines the flavor of the other $B$ meson at the same time. 
The time dependent asymmetry for the decay of the other $B$ meson, the signal $B$, to a CP eigenstates can be written as
\begin{equation}
a_{fCP}(\Delta t) = \frac {\Gamma[\bar B^0(\Delta t)] - \Gamma[B^0(\Delta t)]} {\Gamma[\bar B^0(\Delta t)] + \Gamma[B^0(\Delta t)]} =  S \sin(\Delta m \Delta t) - C \cos(\Delta m \Delta t), 
\label{asymmetry}
\end{equation}
where $\Delta t$ is the difference in decay time between the flavor of the $B$ meson is observed and the time where the meson decays into a CP eigenstate 
and $\Delta m$ is the mass difference. The difference in decay time can be calculated measuring the decay length through $\Delta t = \Delta z/ \beta \gamma c$. 
The coefficients $S$ and $C$ which appear in \ref{asymmetry} are proportional to different types of CP violation: $C$ accounts for ``direct'' CP violation 
while $S$ accounts for the ``mixing induced'' CP violation. For $b \rightarrow c \bar c s$ and $b \rightarrow q \bar q s$ transitions $C \simeq 0$ and $S = -\xi_f \sin 2 \phi_1$. 
In this kind of measurement the main source of systematic uncertainty comes from the resolution on the vertex of the tagging $B$ meson and the performance of the flavor tagging algorithm. 
For the benchmark channels $B \rightarrow J/\psi K_S$ thanks to the improved vertex detector, 
even with the reduced CM boost of Belle II with respect to Belle, the resolution on $\Delta t$ is expected to improve from 0.92 ps to 0.77 ps. 
The simulation also shows that the new flavor tagging algorithm has an efficiency of 35.8\% for Belle II which would represent a 20\% improvement over Belle. 
Table \ref{tab_ccs} shows the comparison of the sensitivity for $B \rightarrow J/\psi K_S$ and for the sum over all  $b \rightarrow c \bar c s$ modes for 
Belle with 1 ab$^{-1}$ and Belle II with 50 ab$^{-1}$ along with the values measured by Belle. The systematic error consists of two parts: a part which is reducible increasing statistics and a irreducible part. 
We quote the irreducible part in two different scenarios: one in which there is no improvement with respect to Belle and with the improvements expected due to better vertexing of Belle II. 
\begin{table}[!htbp]
\centering
%\caption{Comparison of percentages.}
\begin{tabular}{|c|c|c|c|c|c|c|c|}
\hline
\multicolumn{2}{|c}{}  &  \multicolumn{6}{|c|}{Belle (1 ab$^{-1}$)}\\
\hline
Decay Mode & Quantity  & Value & Stat. ($\times 10^{-3}$) & \multicolumn{2}{c|}{Syst. (1) ($\times 10^{-3}$)} & \multicolumn{2}{c|}{Syst. (2) ($\times 10^{-3}$)}\\
\hline
\multicolumn{4}{|c|}{}  & Red. & Non-red. & Red. & Non-red.\\
\hline
\hline
\multirow{2}{*}{$B \rightarrow J/\psi K_S$}& $S $ & +0.67 & 29   & - & 13 & - & {-}\\
    			  & $ C$ & -0.015 & 21 & - & +45, -23 & - & {-}\\	
\hline
\hline
\multirow{2}{*}{$b \rightarrow c \bar c s$}& $S $ & +0.667 & 23 & - & 12 & {-} & -\\
    			  & $ C$ & +0.006 & 16 & - &12 & {-} & -\\	
\hline
\hline
\hline
\multicolumn{2}{|c}{}  & \multicolumn{6}{|c|}{Belle II (50 ab$^{-1}$)}\\
\hline
\multirow{2}{*}{$B \rightarrow J/\psi K_S$}& $ S $ & -  & 3.5 & 1.2 & 8.3  & 1.2 & 4.4\\
    			  & $ C$ & {-}  & 2.5 & 0.7 & +43, -22  & 0.7 & +42, -11\\	
\hline
\hline
\multirow{2}{*}{$b \rightarrow c \bar c s$}& $ S $ & {-}  &2.7 & 2.6 & 7 & 2.6 & 3.6\\
    			  & $C$ & {-}  & 1.9 & 1.4 & 10.6 & 1.4 & 8.7\\
\hline
\end{tabular}
\caption{Comparison of the sensitivity for $S$ and $C$ for the benchmark channel $B \rightarrow J/\psi K_S$ and for the sum over all  $b \rightarrow c \bar c s$ modes for Belle and Belle II. 
For the irreducible systematic error at Belle II we compare two scenarios: one with no major improvement with respect to Belle (Syst. 1) and one with the expected improvements due to the better vertex detector (Syst. 2) (see the text for details). 
For the Belle measurement refer to \cite{physics}.}
\label{tab_ccs}
\end{table}
We studied two different $B$ decays involving $b \rightarrow q \bar q s$ transitions: $B \rightarrow \phi K^0$ and $B \rightarrow \eta' K^0$. 
$B \rightarrow \phi K^0$ is reconstructed in $\phi \rightarrow K^+ K^-$ and $\phi \rightarrow \pi^+ \pi^- \pi^0$ modes with $K_S \rightarrow \pi^+ \pi^-$ and $K_S \rightarrow \pi^0 \pi^0$ (only for $\phi \rightarrow K^+ K^-$). 
The corresponding expected sensitivity is shown in table \ref{phiK}. 
\begin{table}[!hbtp]
\centering
\begin{tabular}{|l|c|c|}
\hline
Decay Mode & $\sigma(S)$ & $\sigma(C)$\\
\hline
\hline
$\phi (K^+ K^-) K_S (\pi^+ \pi^-)$ &     0.025     & 0.017\\
\hline
$\phi (K^+ K^-) K_S (\pi^0 \pi^0)$ & 0.042 & 0.030\\
\hline
$\phi (\pi^+ \pi^- \pi^0) K_S (\pi^+ \pi^-)$ & 0.048 & 0.036\\
\hline
\hline
$K_S$ modes inclusive & 0.019 & 0.014\\
\hline
$K_S$ and $K_L$ modes inclusive & 0.015 & 0.011\\
\hline
\end{tabular}
\caption{Comparison of the sensitivity for the parameters $S$ and $C$ for $B \rightarrow \phi K^0$ for different final states and for inclusive final states at Belle II with 50 ab$^{-1}$}
\label{phiK}
\end{table}
For $B \rightarrow \phi K^0$ and $B \rightarrow \eta' K^0$ where $\eta' \rightarrow \eta \pi^+ \pi^-$ the main difficulty is the reconstruction of the $\pi^0$s and the $\eta$ itself. 
Mis-reconstruction of the neutral candidates leads to signal cross-feed and potentially to a systematically limited resolution. 
Belle II expectations with 50 ab$^{-1}$ are shown in \ref{etaKsens}. 
%Table \ref{etaKres} summarizes the effect of signal cross-feed on the total resolution for $\eta \rightarrow \gamma \gamma$ and $\eta \rightarrow \pi^0 \pi^0 \pi^0$ final states. 
%\begin{table}[!hbtp]
%\centering
%\begin{tabular}{|l|c|c|c|}
%\hline
%Decay Mode & True (ps) & Cross-feed (ps) & Total (ps)\\
%\hline
%\hline
%$\eta' (\eta_{\gamma \gamma} \pi^+ \pi^-) K_S (\pi^+ \pi^-)$ & 1.22 & 2.87 & 1.45\\
%\hline
%$\eta' (\eta_{3 \pi} \pi^+ \pi^-) K_S (\pi^+ \pi^-)$ & 1.17 & 2.36 & 1.50\\
%\hline
%\end{tabular}
%\label{etaKres}
%\caption{$\Delta t$ resolution, in pico-seconds, for $\eta \rightarrow \gamma \gamma$ and $\eta \rightarrow \pi^0 \pi^0 \pi^0$ final states for correctly reconstructed decays, signal cross-feed and the full sample.}
%\end{table}
\begin{table}[!hbtp]
\centering
\begin{tabular}{|l|c|c|}
\hline
Decay Mode & $\sigma(S)$ & $\sigma(C)$\\
\hline
\hline
$\eta' (\eta_{\gamma \gamma} \pi^+ \pi^-) K_S (\pi^+ \pi^-)$ &     0.019     & 0.013\\
\hline
$\eta' (\eta_{3 \pi} \pi^+ \pi^-) K_S (\pi^+ \pi^-)$ & 0.035 & 0.025\\
\hline
\hline
$K_S$ modes inclusive & 0.009 & 0.007\\
\hline
$K_L$ modes inclusive & 0.025 & 0.016\\
\hline
$K_S$ and $K_L$ modes inclusive & 0.0085 & 0.0063\\
\hline
\end{tabular}
\caption{Comparison of the sensitivity for the parameters $S$ and $C$ for $B \rightarrow \eta' K^0$ for different final states and for inclusive final states at Belle II with 50 ab$^{-1}$}
\label{etaKsens}
\end{table}

\section{Measurement of $\phi_2$}
\label{sec_phi2}
To extract $\phi_2$ from $B$ decays we apply an analysis strategy similar to the one used to extract $\phi_1$ to $B \rightarrow h h$ decays, where $h=\pi$, $\rho$. 
\begin{figure}[htb!]
\centering
\includegraphics[width=0.7\textwidth, angle=0]{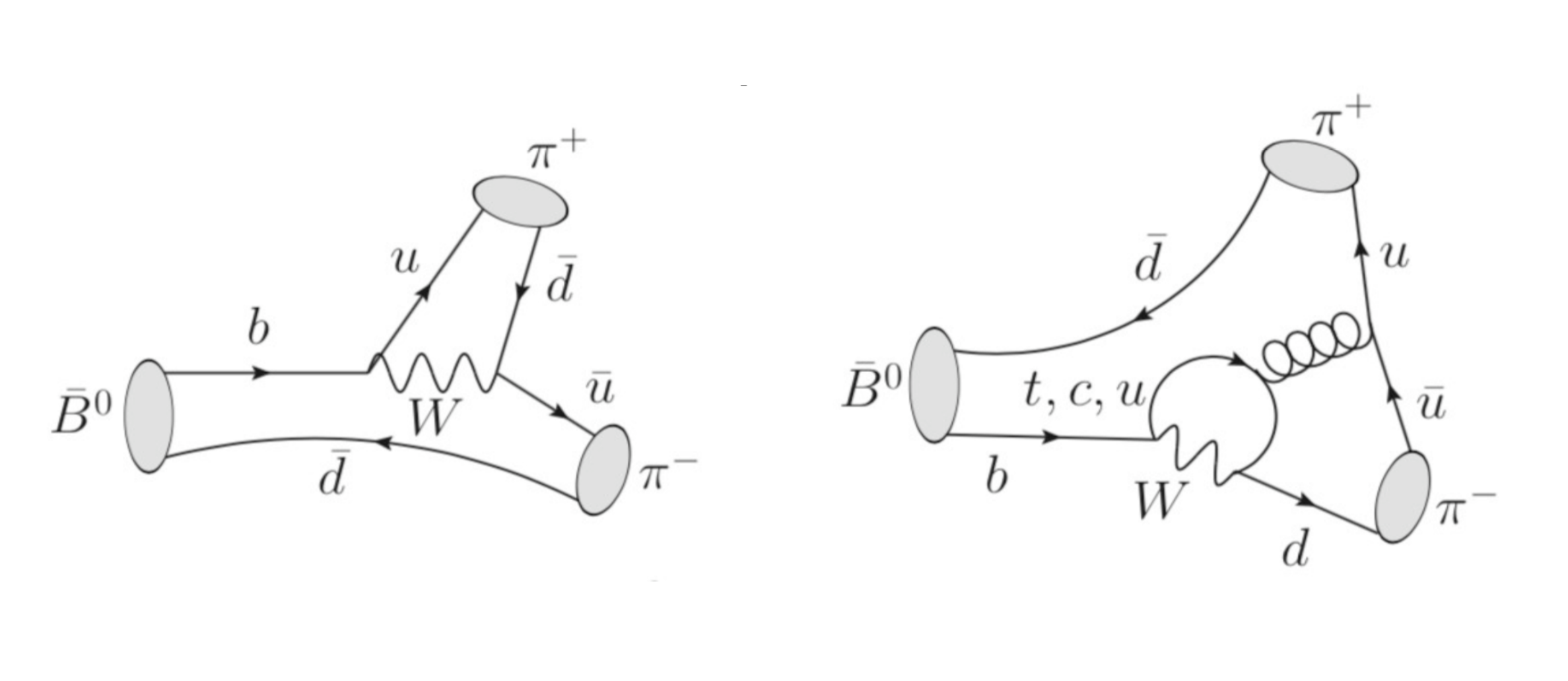} %[width=8.2cm, clip]
\caption{Leading contributions to $B \rightarrow \pi^+ \pi^-$ decays.}
\label{isodiagrams}
\end{figure} 
Differently from $B \rightarrow c \bar c s$ and $B \rightarrow q \bar q s$, in which a single diagram gives the leading contribution, in $B \rightarrow h h$ 
the tree and penguin diagrams shown in Fig. \ref{isodiagrams} have comparable size. 
The different weak and strong phases lead to a shift in $S$ of the form $S=-\xi_f \sin \phi_{2, eff}$ where $\phi_{2, eff} = \phi_2 + \delta \phi$. 
Isospin analysis \cite{gronau} offers the best way to disentangle the contribution of $\delta \phi$. To perform isospin analysis 
one has to measure separately the decays $B^{(i+j)} \rightarrow h^i h^j$ where $h=\pi$, $\rho$ and $i$, $j=\pm$, $0$. 
Up to date $S_{ \pi^0 \pi^0}$ has never been measured and this leaves an eightfold ambiguity on $\phi_2$ in the $B \rightarrow \pi \pi$ channel. %as shown in Fig. \ref{alpha}. 
%\begin {figure}[ht]
%\begin{center}
%\centering
%\begin{minipage}[b]{0.43\textwidth}
%\includegraphics[width=\textwidth]{alpha.pdf}
%\end{minipage}
%\hfill
%\begin{minipage}[b]{0.43\textwidth}
%\includegraphics[width=\textwidth]{alpha_full.pdf}
%\end{minipage}
%\caption{Comparison of the sensitivity for $\phi_2$ for Belle and Belle II without (left) the measurement of $S_{\pi^0 \pi^0}$ and resolution of the ambiguity including the measurement of $S_{ \pi^0 \pi^0}$ at Belle II (right).}
%\label{alpha}
%\end{center}
%\end{figure}
The main difficulty in measuring $S_{ \pi^0 \pi^0}$ lies in the fact that to reconstruct the decay vertex one actually needs tracks; for $B \rightarrow \pi^0 \pi^0$ this translates in 
the request that either one of the photons from the $\pi^0$ decay converts in the inner tracking detectors or that the the $\pi^0$ undergoes a Dalitz decay $\pi^0 \rightarrow e^+ e^- \gamma$. 
According to Belle II simulation the number of $B \rightarrow \pi^0 \pi^0$ decays in which 
one of the two previous conditions holds represents $\simeq 5\%$ of all $B \rightarrow \pi^0 \pi^0$ decays and hence very good reconstruction efficiency along with high statistics is needed. 
Table \ref{alphapipi} lists Belle results and sensitivity and the expected sensitivity for Belle II for charged and neutral final states. 
\begin{table}[!hbtp]
\centering
\begin{tabular}{|c|c|c|c|}
\hline
Quantity & Value & Belle (0.8 ab$^{-1}$) & Belle II (50 ab$^{-1}$)\\
\hline
\hline
B($B \rightarrow \pi^+ \pi^-$) $(\times 10^{-6})$ & $5.04$ & $\pm 0.21 \pm 0.18$ & $\pm 0.03 \pm 0.08$\\
\hline
B($B \rightarrow \pi^0 \pi^0$) $(\times 10^{-6})$ & $1.31$ & $\pm 0.19 \pm 0.18$ & $\pm 0.04 \pm 0.04$\\
\hline
B($B \rightarrow \pi^+ \pi^0$) $(\times 10^{-6})$ & 5.86 & $\pm 0.26 \pm 0.38$ & $\pm 0.04 \pm 0.09$\\
\hline
\hline
$C_{\pi^+ \pi^-}$ & $-0.33$ & $\pm 0.06 \pm 0.03$ & $\pm 0.01 \pm 0.03$\\
\hline
$S_{\pi^+ \pi^-}$ & $-0.64$ & $\pm 0.08 \pm 0.03$ & $\pm 0.01 \pm 0.01$\\
\hline
$C_{\pi^0 \pi^0}$ & $-0.14$ & $\pm 0.36 \pm 0.12$ & $\pm 0.03 \pm 0.01$\\
\hline
$S_{\pi^0 \pi^0}$ & - & - & 1.50\\
\hline
\end{tabular}
\caption{Comparison of the sensitivity for the branching fractions and the parameters $S$ and $C$ for $B \rightarrow \pi \pi$ for charged and neutral final states in Belle and Belle II. For the Belle measurement refer to \cite{alphapipi3}-\cite{alphapipi1}.}
\label{alphapipi}
\end{table}
For $B \rightarrow \rho \rho$ thanks to the presence of at least two charged pions in the final state for both charged and neutral modes 
the full set $C_{\rho^+ \rho^-}$, $S_{\rho^+ \rho^-}$, $C_{\rho^0 \rho^0}$ and $S_{\rho^0 \rho^0}$ has been measured. 
Current measurements along with the expectations for Belle II are reported in table \ref{alpharhorho}. 
\begin{table}[!hbtp]
\centering
\begin{tabular}{|c|c|c|c|}
\hline
Quantity & Value & Belle (0.8 ab$^{-1}$) & Belle II (50 ab$^{-1}$)\\
\hline
\hline
$f_{L, \mbox{  } \rho^+ \rho^-}$ & $0.988$ & $\pm0.012\pm0.023$ & $\pm0.002\pm0.003$\\
\hline
$f_{L, \mbox{  } \rho^0 \rho^0}$ & $0.21$ & $\pm0.20\pm0.15$ & $\pm0.03\pm0.02$\\
\hline
$f_{L, \mbox{  } \rho^+ \rho^0}$ & $0.95$ & $\pm0.11\pm0.02$ & $\pm0.004\pm0.003$\\
\hline
\hline
B($B \rightarrow \rho^+ \rho^-$) $(\times 10^{-6})$ & $28.3$ & $\pm 1.5 \pm 1.5$ & $\pm 0.19 \pm 0.04$\\
\hline
B($B \rightarrow \rho^0 \rho^0$) $(\times 10^{-6})$ & $1.02$ & $\pm 0.30 \pm 0.15$ & $\pm 0.04 \pm 0.02$\\
\hline
B($B \rightarrow \rho^+ \rho^0$) $(\times 10^{-6})$ & 31.7 & $\pm 7.1 \pm 5.3$ & $\pm 0.3 \pm 0.5$\\
\hline
\hline
$C_{\rho^+ \rho^-}$ & $0.0$ & $\pm 0.10 \pm 0.06$ & $\pm 0.01 \pm 0.01$\\
\hline
$S_{\rho^+ \rho^-}$ & $-0.13$ & $\pm 0.15 \pm 0.05$ & $\pm 0.02 \pm 0.01$\\
\hline
$C_{\rho^0 \rho^0}$ & $0.2$ & $\pm 0.8 \pm 0.3$ & $\pm 0.08 \pm 0.01$\\
\hline
$S_{\rho^0 \rho^0}$ & $0.3$ & $\pm 0.7 \pm 0.2$ & $\pm 0.07 \pm 0.01$\\
\hline
\end{tabular}
\caption{Comparison of the sensitivity for the polarization, the branching fractions and the parameters $S$ and $C$ 
for $B \rightarrow \rho \rho$ for both charged and neutral final states at Belle and Belle II. For the Belle measurement refer to \cite{alpharhorho1}-\cite{alpharhorho3}.}
\label{alpharhorho}
\end{table}

\section{Measurement of $\phi_3$}
\label{sec_phi3}
The third angle of the CKM matrix $\phi_3$ is the phase between $b \rightarrow u$ and $b \rightarrow c$. The best methods to measure $\phi_3$ are based on the interference between 
$b \rightarrow c \bar u s$ and $b \rightarrow u \bar c s$ amplitudes with $D^0/ \bar D^0$ decaying to same final state. 
While these transitions are theoretically very clean, they are experimentally very challenging because of CKM and color suppression. 
The first Belle II sensitivity evaluation has been performed using the Dalitz-plot analysis of self-conjugate $D$ decays (GGSZ) in $B^\pm \rightarrow D^0 K^\pm$, $D^0 \rightarrow K_S \pi^+ \pi^-$. 
This analysis strategy requires the knowledge of the different strong phases which have to be measured separately at a charm factory. 
The expected sensitivity for this analysis alone at Belle II with 50 ab$^{-1}$ is $\delta \phi_3 \simeq 3\degree$. 
Combining the GGSZ analysis with other analysis techniques previously used at Belle (ADS, GLW) the combined expected sensitivity is $\delta \phi_3 \simeq 1.6\degree$.

\section{Outlook}
The huge dataset of 50 ab$^{-1}$ of $e^+ e^-$ collisions which is expected to be recorded by the Belle II detector in 5 years of data taking starting in 2019 
together with substantial improvements in the detector, will allow to test the CKM paradigm at the \% level. 
The parameter $\sin(2 \phi_1)$ is expected to be measured with $\simeq$1\% a precision. 
The parameter $\sin(2 \phi_2)$ will benefit from improved isospin analysis that will reduce ambiguities in the $\phi_2$ solution. 
The precision on $\phi_3$ will likely reach $1.6\degree$. 

\begin {figure}[ht]
\begin{center}
%  \centering
%  \begin{minipage}[b]{0.756\textwidth}
%  \includegraphics[width=\textwidth]{current_ut.pdf}
% \end{minipage}
 % \hfill
  \begin{minipage}[b]{0.758\textwidth}
\includegraphics[width=\textwidth]{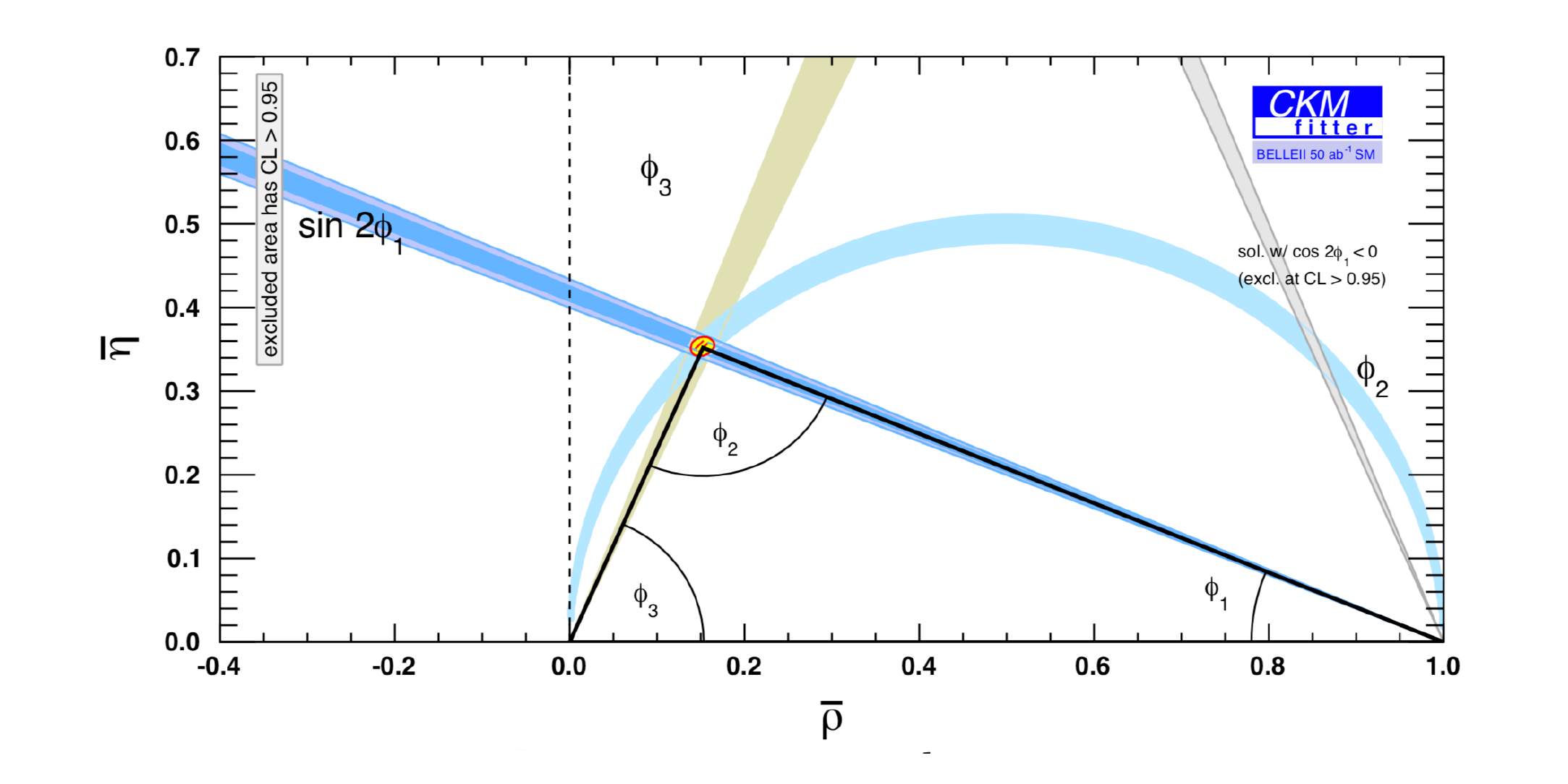}
  \end{minipage}
\caption{Bounds on the Unitarity Triangle after Belle II will have recorded 50 ab$^{-1}$ using only the measurements discussed in this article.}
     \label{ut}
\end{center}
\end{figure}

\end{document}